\documentclass[prd,nofootinbib,superscriptaddress,preprint]{revtex4}
\usepackage[T1]{fontenc}
\usepackage{amsmath,amssymb}
\usepackage{epsfig}
\usepackage{graphicx}
\usepackage[usenames,dvipsnames]{color}
\usepackage{slashed}
\usepackage[colorlinks,citecolor=blue]{hyperref}
\usepackage{pdfpages}
\usepackage{color}
\usepackage{subfig}

\newcommand{\R}{\mbox{\tiny$R$}}
\newcommand{\T}{\mbox{\tiny$T$}}
\newcommand{\s}{\mbox{\tiny$S$}}

\begin{document}

\title{\Large Linear Seesaw for Dirac Neutrinos \\ with $A_4$ Flavour Symmetry} 
\author{Debasish Borah}
\email{dborah@iitg.ac.in}
\affiliation{Department of Physics, Indian Institute of Technology
Guwahati, Assam 781039, India}
\author{Biswajit Karmakar}
\email{biswajit@prl.res.in}
\affiliation{Theoretical Physics Division, Physical Research Laboratory, Ahmedabad 380009, India}

\begin{abstract}
We propose a linear seesaw model to realise light Dirac neutrinos within the framework of $A_4$ discrete flavour  symmetry. The
additional fields and their transformations under the flavour  symmetries are chosen in such a way that naturally predicts the
hierarchies of  different elements of the seesaw mass matrix and also keeps the unwanted terms away. For generic choices of
flavon alignments, the model predicts  normal hierarchical light neutrino masses with the atmospheric mixing angle in  the lower
octant. Apart from predicting interesting correlations between  different neutrino parameters as well as between neutrino and 
model parameters,  the model also predicts the leptonic Dirac CP phase to lie in a specific range  $-\pi/2\lesssim \delta 
\lesssim -\pi/5$ and  $\pi/5\lesssim \delta \lesssim \pi/2$ that includes the currently preferred maximal value. The predictions
for the absolute neutrino  masses in one specific version of the model can also saturate the cosmological upper bound on sum of
absolute neutrino masses.
\end{abstract}

\maketitle 

\section{Introduction}
The fact that neutrinos have non-zero but tiny masses, several order of magnitudes smaller compared to the electroweak scale and 
large mixing \cite{Patrignani:2016xqp} has been verified again and again in the last two decades. The present status of neutrino
oscillation data can be found in the recent global fit analysis \cite{Esteban:2016qun, deSalas:2017kay, nufit}, which clearly 
indicate that we do not yet know some of the neutrino parameters namely, the mass hierarchy of neutrinos: normal $(m_3 >  m_2 >
m_1)$ or inverted $(m_2 > m_1 > m_3)$, leptonic CP violation as   well as the octant of atmospheric mixing angle $\theta_{23}$.
While the next generation neutrino experiments will be able to settle these issues, one still can not determine the nature of 
neutrino: Dirac or Majorana in neutrino oscillation experiments. Though Majorana nature of neutrinos can be probed through lepton
number violating signatures like neutrinoless double beta decay $(0\nu\beta\beta)$, there has not been any positive signal of it
yet. For example, please refer to the latest results from KamLAND-ZEN experiment \cite{KamLAND-Zen:2016pfg}. Although such null 
results only disfavour the quasi-degenerate regime of light Majorana neutrinos and can never rule out Majorana nature of 
neutrinos, this has recently motivated the particle physics community to study the scenario of Dirac neutrinos with similar 
interest as given to Majorana neutrinos in the last few decades. The conventional seesaw mechanism for the origin of neutrino 
masses \cite{Minkowski:1977sc, GellMann:1980vs, Mohapatra:1979ia, Schechter:1980gr} and its many descendants predict light 
Majorana neutrinos. On the contrary, there were fewer proposals to generate light Dirac neutrino masses initially~\cite{Babu:1988yq,
Peltoniemi:1992ss} but it has recently gained momentum with several new proposals to realise sub-eV scale Dirac neutrino 
masses~\cite{Chulia:2016ngi, Aranda:2013gga, 
Chen:2015jta, Ma:2015mjd, Reig:2016ewy, Wang:2016lve, Wang:2017mcy, Wang:2006jy, 
Gabriel:2006ns, Davidson:2009ha, Davidson:2010sf, Bonilla:2016zef, 
Farzan:2012sa, Bonilla:2016diq, Ma:2016mwh, Ma:2017kgb, Borah:2016lrl, 
Borah:2016zbd, Borah:2016hqn, Borah:2017leo, CentellesChulia:2017koy, 
Bonilla:2017ekt, Memenga:2013vc, Borah:2017dmk, CentellesChulia:2018gwr, 
CentellesChulia:2018bkz, Han:2018zcn, Borah:2018gjk}. Since the coupling of left and right handed neutrinos to the standard model
(SM) Higgs field will require fine tuning of Yukawa coupling to the level of $10^{-12}$ or even less, it is important to forbid 
such couplings at tree level by introducing some additional symmetries such as $U(1)_{B-L}, Z_N, A_4$ which also make sure that
the right handed singlet neutrinos do not acquire any Majorana mass terms.

There have been several discussions on other conventional seesaw mechanisms in the context of Dirac neutrinos for example, type I
seesaw~\cite{CentellesChulia:2017koy, Borah:2017dmk}, type II seesaw~\cite{Bonilla:2017ekt}, inverse seesaw~\cite{Borah:2017dmk}
and so on. Here show how light Dirac neutrinos can be realised within another seesaw scenario, known as linear seesaw mechanism. 
We consider the presence of $A_4$ flavour symmetry augmented by additional discrete $Z_N$ and global lepton number symmetries
which not only dictate the neutrino mixing patterns but also keep the unwanted terms away from the seesaw mass matrix, in order
to realise linear seesaw. Linear seesaw for Majorana neutrino was proposed in earlier works~\cite{Malinsky:2005bi,Hirsch:2009mx}
and further extended to radiative seesaw models in~\cite{Wang:2015saa, Das:2017ski} and hidden gauge sector models 
in~\cite{Nomura:2018ibs}. We extend it to Dirac neutrino scenarios in a minimal way incorporating the above-mentioned flavour 
symmetries. Apart from retaining the usual attractive feature of linear seesaw, like the viability of seesaw scale at TeV
naturally without much fine-tuning, the model also predicts several other aspects of neutrinos that can be tested at upcoming 
experiments. Among them, the preference for normal hierarchy, specific range of Dirac CP phase that includes the maximal value,
atmospheric mixing angle in lower octant are the ones which address the present puzzles in neutrino physics.

Rest of the paper is organised as follows. In section \ref{sec1}, we discuss the conventional and Dirac linear seesaw model
and its predictions for sub-eV Dirac neutrinos in details. Finally, we conclude in section \ref{sec2}.

\section{The Linear Seesaw Model}
\label{sec1}
In the conventional linear seesaw model for Majorana neutrinos~\cite{Malinsky:2005bi,Hirsch:2009mx}, the standard model fermion 
content is effectively extended by two different types of neutral singlet fermions $(N, S)$ per generation and the complete 
neutral fermion mass matrix ($9\times9$) in the basis ($\nu_L, N, S$) assumes the form 
\begin{equation}\label{eq:1}
      M_{\nu}= \left(\begin{array}{ccc}
      0 & m_{D} & M_L \\
      m_{D}^{T}& 0 & M\\ 
      M_L^{T} & M^{T} & 0
      \end{array}\right), 
\end{equation}
due to the chosen symmetries or scalar content of the model. The light neutrino mass matrix can be derived from this as
       \begin{equation}\label{eq:lss1} 
        m_{\nu} = m_{D}(M_L M^{-1})^T+(M_L M^{-1})m_{D}^T,  
       \end{equation}
which, being linear in Dirac neutrino mass matrix $m_D$ is known as the linear  seesaw~\cite{Malinsky:2005bi,Hirsch:2009mx}. Here
the effective light neutrino mass is roughly given by $\sim \epsilon m_{D}/M$ where  $\epsilon$ is originated from a small lepton
number violating term in $M_L$, the (13) entry of the neutral fermion mass matrix given in Eq. \eqref{eq:1}. This is a simple 
alternative to the usual inverse seesaw model~\cite{Mohapatra:1986aw, Mohapatra:1986bd, GonzalezGarcia:1988rw, Catano:2012kw} 
where we also introduce  two sets of gauge singlet Majorana neutrinos at the TeV scale to obtain light  neutrino mass in sub-eV 
range. We now consider an extension of this simple linear seesaw model to generate sub-eV Dirac neutrino masses following a 
similar roadmap that was used to accommodate light Dirac neutrinos in type I seesaw~\cite{CentellesChulia:2017koy, Borah:2017dmk},
type II seesaw~\cite{Bonilla:2017ekt}, inverse seesaw~\cite{Borah:2017dmk} etc. Apart from introducing the right handed 
counterpart of the usual left handed neutrinos, the other heavy fermions introduced for seesaw purpose are of Dirac nature,
having both helicities: ($N_L, N_R$) and ($S_L, S_R$). In such a case the complete linear seesaw mass matrix can be written in 
$(\nu_L, N_L, S_L)^T, (\nu_R, N_R, S_R)$ basis as
\begin{equation}\label{eq:1a}
      m_{\nu}= \left(\begin{array}{ccc}
      0 & m_{\nu N} & M_{\nu S} \\
      m_{\nu N}^{\prime}& 0 & M_{NS}\\ 
      M^{\prime}_{\nu S} & M^{\prime}_{NS} & 0
      \end{array}\right). 
\end{equation}
The corresponding formula for light Dirac neutrinos can be written as
\begin{eqnarray}\label{eq:nulss}
        m_{\nu} 
                &=& m_{\nu N}(M'_{\nu S} M_{NS}'^{-1})+(M_{\nu S} 
M_{NS}^{-1})m'_{\nu N}. 
\end{eqnarray}
Similar to the linear seesaw for Majorana neutrinos, here also if we demand the two terms $m'_{\nu N}, M'_{\nu S}$ to be small, one can generate light Dirac neutrino masses in a TeV scale linear seesaw mechanism. However, the naturalness of such small terms can not be due to an approximate lepton number symmetry, as we are considering exact global lepton number symmetry to have purely Dirac nature of light neutrinos. Therefore, we can consider another global symmetry which is approximately broken by the two terms $m'_{\nu N}, M'_{\nu S}$ which naturally makes them small according to the t'Hooft's naturalness criterion \cite{tHooft:1979rat}. It is worth noting that both these terms contain the right handed neutrino $\nu_R$. Therefore under the approximate global symmetry $U(1)_X$, say, the singlet fermion $\nu_R$ can have some non-zero charge so that the two terms $m'_{\nu N}, M'_{\nu S}$ explicitly break this symmetry, providing a natural origin of their smallness.

Here we intend to present an $A_4$ flavour symmetric model of Dirac linear seesaw, in a way similar to the Altarelli-Feruglio model \cite{Altarelli:2005yx} for Majorana light neutrinos. Since minimal versions of such models often involve higher dimensional operators to allow coupling of $A_4$ flavons (which are SM singlets) and leptons, we first briefly discuss a renormalisable version of Dirac linear seesaw before discussing an $A_4$ symmetric one in details. In order to generate the desired structure we also choose discrete symmetries $Z_4 \times Z_3$. The presence of an approximate global symmetry $U(1)_X$ is also assumed in order to naturally accommodate a TeV scale version of such seesaw. The field content and the transformations under the chosen symmetries are shown in table \ref{tab:lss1}. The renormalisable Yukawa Lagrangian can be written as
\begin{align}
\mathcal{L}_Y & \supset \sum_{\alpha, \beta} Y_e \bar{L}_{\alpha} H e_{\beta} + Y_2 \bar{L} \tilde{H} N_R + Y_3 \bar{L} \tilde{H} S_R + Y_4 \bar{\nu_R} N_L \phi_1 \nonumber \\
& + Y_5 \bar{S_L} \nu_R \phi^{\dagger}_1 + Y_6 \bar{S_R} N_L \phi_2 + Y_7 \bar{S_L} N_R \phi^{\dagger}_2 +{\rm h.c.}
\end{align}
Therefore, after the SM Higgs and the singlet scalar fields acquire non-zero vacuum expectation value (vev), one can generate the Dirac linear seesaw mass matrix \eqref{eq:1a} with different terms given by 
$$ m_{\nu N} = Y_2 \langle H \rangle, M_{\nu S} =Y_3 \langle H \rangle, m'_{\nu N}= Y_4 \langle \phi_1 \rangle, $$
$$ M'_{\nu S} = Y_5 \langle \phi_1 \rangle, M_{NS}=Y_6 \langle \phi_2 \rangle, M'_{NS}=Y_7 \langle \phi_2 \rangle.$$
Now, under the approximate global symmetry $U(1)_X$ mentioned above, we can have either $\nu_R$ or the singlet scalar $\phi_1$ having non-trivial charge. In the case of approximate global symmetry $U(1)_X$ under which only $\nu_R$ is charged, one can have the Yukawa couplings $Y_4, Y_5$ naturally small as $Y_{4,5} \rightarrow 0$ helps us recover the full global $U(1)_X$ symmetry. This also makes the terms $m'_{\nu N}, M'_{\nu S}$ naturally small, required for the realisation of a TeV scale linear seesaw. On the other hand, if $\phi_1$ is charged under $U(1)_X$ and its mass squared term is positive definite, then it can acquire an induced vev from soft $U(1)_X$ breaking terms in the scalar potential. However, suitable scalars have to be incorporated to allow such soft $U(1)_X$ breaking terms that do not break any of the other symmetries.

The charged lepton mass is given by $m_l = Y_e \langle H \rangle $. We also note that both $Z_4$ and $Z_3$ are required 
for the desired structure of the linear seesaw mass matrix. For example, if we do not have $Z_3$ symmetry, then for the
given $Z_4$ charges, we can also write terms like $\bar{N_L} N_R \phi_2, \bar{S_L} S_R \phi_2$ which will give non-zero
contribution to the (22) and (33) entries of the linear seesaw mass matrix \eqref{eq:1a}. Therefore, we retain both the
discrete symmetries. Since there is no $A_4$ or similar non-abelian discrete symmetries acting over the generation of
fermions, this model does not further predict the structure of different Yukawa coupling mass matrices involved. So 
there exists lots of freedom in fitting the resulting light neutrino mass matrix with the neutrino oscillation data. To
make the model more predictive, we now consider an $A_4$ flavour symmetric version of Dirac linear seesaw and numerically
analyse its predictions for the light neutrino sector. The price we have to pay for such minimality is to go beyond the
renormalisable level thereby introducing dimension five operators. We also utilise these higher dimensional operators to
explain the hierarchy between different terms of the linear seesaw mass matrix and hence do not have the requirement of
any approximate global symmetry like $U(1)_X$ discussed above. It should be noted that the renormalisable
versions of such seesaw mechanisms, in principle, do not necessarily require exact or approximate global symmetries
like $U(1)_L, U(1)_X$. As shown in \cite{Hirsch:2017col}, the global lepton number symmetry broken to an appropriate
$Z_n$ subgroup is enough to  protect Dirac nature of neutrinos and invoke naturalness arguments justifying linear Dirac
seesaw discussed above.
\begin{table}[h]
\centering
\resizebox{10cm}{2cm}{%
\begin{tabular}{|c|cccccccc|cc|}
\hline
 Fields & $L$  & $e_{\R}, \mu_{\R}, \tau_{\R}$ &  $H$ & $\nu_{\R}$& $N_L$ & 
$N_{\R}$ & $S_L$ & $S_R$ & $\phi_1$ & $\phi_2$ \\ \hline
$SU(2)_L$ & $2$ &$1$ & 2 & 1& $1$  & $1$& $1$ & $1$ & $1$ & 1
\\
\hline
$Z_{4}$ & $i$ &$i$ & 1 & 1& -$i$  & $i$& -$i$ & $i$ & $i$ & -1
\\
\hline
$Z_3$ & 1 & 1 & 1 & $\omega$& $\omega^2$ & 1 & $\omega^2$ & 1 & $\omega^2$ & $\omega$  \\
\hline
\end{tabular}
}\
\caption{\label{tab:lss1}  Fields content and transformation properties under
$Z_4 \times Z_3 $ discrete symmetry for renormalisable Dirac linear seesaw.}
\end{table}

To obtain the desired structure of the seesaw mass matrix given in Eq. \eqref{eq:1a} and to obtain the required hierarchy among 
its elements, we consider $Z_4 \times Z_3$ symmetry and a global lepton number $U(1)_L$ symmetry in addition to $A_4$ flavour 
symmetry which plays a crucial role in realising flavour structures of the corresponding mass matrices.
\begin{table}[h]
\centering
\resizebox{10cm}{2cm}{%
\begin{tabular}{| c|cccccccc|ccccc|}
\hline
 Fields & $L$  & $e_{\R}, \mu_{\R}, \tau_{\R}$ &  $H$ & $\nu_{\R}$& $N_L$ & 
$N_{\R}$ & $S_L$ & $S_R$ & $\phi_{\s}$ & $\phi_{\T}$ &  $\xi$ & $\eta$ &$\rho$ 
\\ \hline
$SU(2)_L$ & 2 & 1 & 2 & 1 &1 &1& 1 & 1 & 1 & 1 & 1 &1 &1 \\
\hline
$A_{4}$ & 3 & 1,$1''$,$1'$ & 1 & 3 &3 &3& 3 & 3 & 3 & 3 & 1 &1 &1 \\
\hline
$Z_{4}$ & $i$ &$i$ & 1 & 1& -$i$  & $i$& -$i$ & $i$ & -$1$ & 1 & -$1$&-1 & $i$ 
\\
\hline
$Z_3$ & 1 & 1 & 1 & $\omega$& $\omega^2$ & 1 & $\omega^2$ & 1 & $\omega$ & 1& 
$\omega$ &$\omega$ & 1  \\
\hline
$U(1)_L$ & 1 & 1& 0 &1 &1 &1 &1 &1 & 0 & 0 & 0 & 0& 0  \\
\hline
\end{tabular}
}\
\caption{\label{tab:lss2}  Fields content and transformation properties under
$A_4 \times Z_4 \times Z_3 $ symmetry for linear seesaw. }
\end{table}
In table \ref{tab:lss2} we elaborate the transformation of SM  fields as well as the additional fermions and flavons involved in
the present  construction of linear seesaw. Here the SM lepton doublets ($L$) and  the additional singlet neutral fermions 
transform as $A_4$ triplets. On the other hand, SM charged  leptons ($e_{\R}, \mu_{\R}, \tau_{\R}$) transform as 1,$1''$ and
$1'$  respectively under same $A_4$ symmetry. Likewise most $A_4$ flavour models, two $A_4$  triplet flavons present in the 
set-up $\phi_{\T}$  and $\phi_{\s}$, play an  instrumental role in generating the non-diagonal mass matrices for charged 
lepton and neutrino sectors respectively.  The Yukawa Lagrangian upto leading order for charged leptons invariant under this
$A_4\times  Z_4 \times Z_3 \times U(1)_L$ symmetry is
\begin{align}\label{eq:lsslag}
 \mathcal{L}_{\rm CL} &= \frac{y_e}{\Lambda}(\bar{L}\phi_{\T})H e_{\R} +\frac{y_{\mu}}{\Lambda}(\bar{L}\phi_{\T})_{1'}H\mu_{\R}+ 
                                       \frac{y_{\tau}}{\Lambda}(\bar{L}\phi_{\T})_{1''}H\tau_{\R}+\text{h.c.},  
\end{align}
where $\Lambda$ is the cut-off scale of the theory and both $y$'s are the  respective dimensionless Yukawa coupling constants. 
Here the leading order contribution to  the charged leptons via $\bar{L}H \ell_i$ (where $\ell_i$ are the right handed charged 
leptons) are not invariant under $A_4$ symmetry. In presence of the triplet flavon $\phi_{\T}$ one can easily construct $A_4$
invariant dimension five  operators as shown on the right hand side of Eq. \eqref{eq:lsslag} which  subsequently generate the 
relevant masses for charged leptons after flavon $\phi_{\T}$ and the SM  Higgs field acquire non-zero vev. In appendix  \ref{appen1}, we have briefly summarised the $A_4$ product rules which dictate  the flavour structure of 
the mass matrices. Now, for the triplet flavon  $\phi_{\T}$, considering a generic vev alignment $\langle \phi_{\T} \rangle=(v_{\T}, 
v_{\T}, v_{\T})$, the charged lepton mass matrix can be written as  
\begin{eqnarray}\label{mCL2}
m_{l} =\frac{vv_{\T}}{\Lambda} \left(
\begin{array}{ccc}
         y_e & y_{\mu}          & y_{\tau}\\
         y_e & \omega y_{\mu}   & \omega^2 y_{\tau} \\
         y_e & \omega^2 y_{\mu} & \omega y_{\tau} 
\end{array}
\right),
\end{eqnarray}
where $v$ is the vev of the SM Higgs doublet $H$ and $\omega=e^{i2\pi/3}$ is  the cube root of unity. This charged lepton mass
matrix now can be diagonalised  by using a matrix $U_{\omega}$ (also known as the magic matrix), given by
\begin{eqnarray}\label{eq:omega}
U_{\omega} =\frac{1}{\sqrt{3}}\left(
\begin{array}{ccc}
         1 & 1          & 1\\
         1 & \omega     & \omega^2\\
         1 & \omega^2   & \omega 
\end{array}
\right).  
\end{eqnarray}
Now, for neutrino sector the relevant Yukawa Lagrangian is given  by 
\begin{align}\label{eq:lssnu}
 \mathcal{L}_{\nu}=
&~~ Y_{\nu N}\bar{L}\tilde{H}N_R+  \bar{\nu_{\R}}{N_L}
    \left(y'_{\xi_2}\xi^{\dagger}+ y'_{\eta_2}\eta^{\dagger}+y'_{s_2}\phi_s^{\dagger}+ y'_{a_2}\phi_s^{\dagger}\right)\frac{\rho^{\dagger}}{\Lambda}\nonumber   \\ 
&~~~    + Y_{\nu S} \bar{L}\tilde{H}S_R+ \bar{S_L}\nu_{\R}   \left(y'_{\xi_1}\xi+ y'_{\eta_1}\eta+y'_{s_1}\phi_s+ y'_{a_1}\phi_s\right)\frac{\rho}{\Lambda}\nonumber   \\ 
&~~~+ \bar{S_R} N_L (y_{\xi_2}\xi+y_{\eta_2}\eta+ y_{s_2}\phi_s+y_{a_2}\phi_s) \nonumber \\
&~~~+ \bar{S_L} N_R  (y_{\xi_1}\xi^{\dagger}+y_{\eta_1}\eta^{\dagger}+ y_{s_1}\phi^{\dagger}_s+y_{a_1}\phi^{\dagger}_s)+\text{h.c.} 
\end{align}
 Here both $\bar{L}\tilde{H}N_R$ and $\bar{L}\tilde{H}S_R$ terms, involving SM lepton  doublet are generated at tree level.  As 
 both SM  lepton doublets ($L$) and gauge singlet Dirac fermions ($N, S$) are  $A_4$ triplets (the SM Higgs $H$ being a singlet
 under the same), following the $A_4$ multiplication rules given in appendix \ref{appen1}, we find the associated  mass matrices
 to be diagonal. These mass matrices can be written  as  
\begin{eqnarray}\label{mat:mix1}
 m_{\nu N} = Y_{\nu N}v\mathbf{I},~~M_{\nu S} = Y_{\nu S}v\mathbf{I},  
\end{eqnarray}
where $\mathbf{I}$ is a $3\times 3$ identity matrix. 
On the other hand, owing to the specific discrete $Z_4 \times Z_3$ symmetry,   $S_L$-$\nu_{\R}$ and $\nu_{\R}$-$N_L$ couplings 
are generated at dimension five level,  ensuring the smallness of these couplings. These contributions come via  involvement of 
the $A_4$ singlet flavons $\xi, \eta$ and $\rho$ as well as the triplet flavon  $\phi_{\s}$. Unlike in the conventional linear 
seesaw mechanism for Majorana neutrinos,  here we do not have any approximate global symmetry to make certain terms of the mass 
matrix small, from naturalness arguments. Therefore, we need to assign these additional discrete symmetries so that at least one
of the mass matrices contributing to the light neutrino mass formula in Eq. \eqref{eq:nulss} arises at next to leading order. 
In this set-up $\phi_{\s}$ and $\xi$  share same discrete charges  like $\eta$, hence all of them therefore contribute to the $S_L$-$\nu_{\R}$
and $\nu_{\R}$-$N_L$ couplings. This essentially leads to same non-diagonal  contributions in these couplings. Now, with the vev
alignment for  the flavons $\phi_{\s}$, $\xi$, $\eta$ and $\rho$ as, $\langle \phi_{\s} \rangle=(0, v_{\s},0), \langle 
\xi \rangle=v_{\xi}$, $\langle \eta \rangle=v_{\eta}$ and $ 
\langle \rho \rangle=v_{\rho}$ respectively, the most general mass matrices corresponding to these two couplings  
can be written as 
\begin{eqnarray}\label{mat:lheavy}
 M'_{ \nu S} =\left(
\begin{array}{ccc}
 x'_1 & 0   & s'_1+a'_1 \\
 0  & x'_1 & 0\\
 s'_1 - a'_1 & 0 & x'_1
\end{array}
\right), ~~ m'_{ \nu N} =\left(
\begin{array}{ccc}
 x'_2 & 0   & s'_2+a'_2 \\
 0  & x'_2 & 0\\
 s'_2 - a'_2 & 0 & x'_2
\end{array}
\right), 
\end{eqnarray}
where $x'_1=(y'_{\xi_1} v_{\xi}+y'_{\eta_1}v_{\eta})v_{\rho}/\Lambda$, $s'_1=y'_{s_1} v_{\s} v_{\rho}/\Lambda, a'_1=y'_{a_1} 
v_{\s}v_{\rho}/\Lambda$, $x'_2=(y'_{\xi_2} v_{\xi}+y'_{\eta_2}v_{\eta})v_{\rho}/\Lambda$, $s'_2=y'_{s_2} v_{\s}v_{\rho}/\Lambda$
and $a'_2=y'_{a_2} v_{\s} v_{\rho}/\Lambda$.    Note that $s'_i$ and $a'_i$ (where $i=1,2$) are the symmetric and anti-symmetric contributions 
originated from $A_4$ multiplications. Similarly, the mixing between the heavy neutrinos  $S_L - N_R$ and $S_R- N_L$  are 
generated at dimension four level, in adhesion with the flavons $\phi_{\s}$,   $\xi$, $\eta$.  Now,again  with the vev alignment for 
the flavons $\phi_{\s}$, $\xi$ and $\eta$ as, $\langle \phi_{\s} \rangle=(0, v_{\s},0), \langle  \xi \rangle=v_{\xi},  \langle 
\eta \rangle=v_{\eta}$, the mass matrices involved here can be written as 
\begin{eqnarray}\label{mat:heavy}
 M'_{ N S} =\left(
\begin{array}{ccc}
 x_1 & 0   & s_1+a_1 \\
 0  & x_1 & 0\\
 s_1 - a_1 & 0 & x_1
\end{array}
\right), ~~ M_{ N S} =\left(
\begin{array}{ccc}
 x_2 & 0   & s_2+a_2 \\
 0  & x_2 & 0\\
 s_2 - a_2 & 0 & x_2
\end{array}
\right),
\end{eqnarray}
where $x_1=y_{\xi_1} v_{\xi}+y_{\eta_1} v_{\eta}$, $s_1=y_{s_1} v_{\s}, a_1=y_{a_1}  v_{\s}$, $x_2=y_{\xi_2} v_{\xi}+y_{\eta_2} 
v_{\eta}$, $s_2=y_{s_2} v_{\s}$ and $a_2=y_{a_2}  v_{\s}$. Here also $s_i$ and $a_i$ are the symmetric and anti-symmetric contributions
originated from $A_4$ multiplication. This unique contribution ($a_i$ or $a'_i$)  is a specific  feature of $A_4$ flavour models 
for Dirac neutrinos and usually do not appear  for Majorana neutrinos due to symmetry property of the Majorana mass matrix. 
It is worth mentioning that, these anti-symmetric parts, originated due to the  Dirac nature of neutrinos, significantly dictate 
the pattern of neutrino mixing  and can explain non-zero $\theta_{13}$ in a very minimal scenario~\cite{Memenga:2013vc} compared
to what is usually done with Majorana neutrinos~\cite{Karmakar:2014dva}. Such anti-symmetric contribution  from $A_4$ triplet
products can also play a non-trivial role in generating nonzero  $\theta_{13}$ in Majorana neutrino scenarios (through Dirac 
Yukawa  coupling appearing in type I seesaw) \cite{Borah:2017qdu}. Now, substituting these mass matrices obtained in Eq. 
\eqref{mat:mix1}-\eqref{mat:heavy} in the linear seesaw formula given in Eq. \eqref{eq:nulss} one can obtain   the effective 
light neutrino mass matrix as 
{\small 
\begin{align}
m_{\nu} &= Y_{\nu N}v \left(
\begin{array}{ccc}
 x'_1 & 0   & s'_1+a'_1 \\
 0  & x'_1 & 0\\
 s'_1 - a'_1 & 0 & x'_1
\end{array}
\right)
 \left(
\begin{array}{ccc}
  x_1 & 0   & s_1+a_1 \\
 0  & x_1 & 0\\
 s_1 - a_1 & 0 & x_1
\end{array}
\right)^{-1}\nonumber\\
& ~~~~~~~~~~~~~~+Y_{\nu S}v\left(
\begin{array}{ccc}
 x'_2 & 0   & s'_2+a'_2 \\
 0  & x'_2 & 0\\
 s'_2 - a'_2 & 0 & x'_2
\end{array}
\right)
\left(
\begin{array}{ccc}
 x_2 & 0   & s_2+a_2 \\
 0  & x_2 & 0\\
 s_2 - a_2 & 0 & x_2
\end{array}
\right)^{-1},\nonumber\\
&= \frac{Y_{\nu N}v}{a_1^2-s_1^2+x_1^2} \left(
\begin{array}{ccc}
 (a_1-s_1)(a'_1+s'_1)+x_1x'_1 & 0   & (a'_1+s'_1)x_1-(a_1+s_1)x'_1 \\
 0  & \frac{x'_1(a_1^2-s_1^2+x_1^2)}{x_1} & 0\\
 (a_1-s_1)x'_1-(a'_1-s'_1)x_1 & 0 & (a_1+s_1)(a'_1-s'_1)+x_1x'_1
\end{array}
\right)\nonumber \\
&+ \frac{Y_{\nu S}v}{a_2^2-s_2^2+x_2^2} \left(
\begin{array}{ccc}
 (a_2-s_2)(a'_2+s'_2)+x_2x'_2 & 0   & (a'_2+s'_2)x_2-(a_2+s_2)x'_2 \\
 0  & \frac{x'_2(a_2^2-s_2^2+x_2^2)}{x_2} & 0\\
 (a_2-s_2)x'_2-(a'_2-s'_2)x_2 & 0 & (a_2+s_2)(a'_2-s'_2)+x_2x'_2
\end{array}
\right).
\label{mnu:gen}
\end{align}
}
Here all the elements appearing in the effective neutrino mass matrix are  complex in general. To diagonalise this general
complex matrix let us first define a Hermitian matrix $\mathcal{M}$, given by 
{\small 
\begin{eqnarray}
 \mathcal{M}&=&m_{\nu}m_{\nu}^{\dagger}\nonumber\\
            &=& \left( \label{mat:hgen}
\begin{array}{ccc}
|\lambda_1 p_{1}+\lambda_2 q_1|^2+|\lambda_1 p_{2}+\lambda_2 q_2|^2 & 0   & \begin{smallmatrix}
      (\lambda_1 p_{1}+\lambda_2 q_1)(\lambda_1 p_{4}+\lambda_2 q_4)^* + {} \\
     (\lambda_1 p_{2}+\lambda_2 q_2)(\lambda_1 p_{5}+\lambda_2 q_5)^*
    \end{smallmatrix}\\
 0   & |\lambda_1 p_{3}+\lambda_2 q_3|^2 & 0\\
\begin{smallmatrix}
      (\lambda_1 p_{4}+\lambda_2 q_4)(\lambda_1 p_{1}+\lambda_2 q_1)^* + {} \\
     (\lambda_1 p_{5}+\lambda_2 q_5)(\lambda_1 p_{2}+\lambda_2 q_2)^*
    \end{smallmatrix}& 0 & |\lambda_1 p_{4}+\lambda_2 q_4|^2+|\lambda_1 p_{5}+\lambda_2 q_5|^2
\end{array}
\right)
\end{eqnarray}
}
where 
\begin{align}
 \lambda_1&={Y_{\nu N}v}/({a_1^2-s_1^2+x_1^2}),&\lambda_2&={Y_{\nu S}v}/({a_2^2-s_2^2+x_2^2})\\
 p_1&=(a_1-s_1)(a'_1+s'_1)+x_1x'_1,&q_1&=(a_2-s_2)(a'_2+s'_2)+x_2x'_2,\\
 p_2&=(a'_1+s'_1)x_1-(a_1+s_1)x'_1,&q_2&=(a'_2+s'_2)x_2-(a_2+s_2)x'_2, \\
 p_3&={x'_1(a_1^2-s_1^2+x_1^2)}/{x_1},&q_3&={x'_2(a_2^2-s_2^2+x_2^2)}/{x_2},\\
 p_4&= (a_1-s_1)x'_1-(a'_1-s'_1)x_1,&q_4&=(a_2-s_2)x'_2-(a'_2-s'_2)x_2,\\
 p_5&= (a_1+s_1)(a'_1-s'_1)+x_1x'_1,&q_5&=(a_2+s_2)(a'_2-s'_2)+x_2x'_2. \label{mat:hpara}
\end{align}
This structure of the Hermitian matrix $\mathcal{M}$ suggests that it can be diagonalised by a rotation matrix $U_{13}$ 
satisfying $U_{13}^{\dagger}\mathcal{M}U_{13}={\rm diag}  (m_1^2,m_2^2,m_3^2)$, where 
\begin{eqnarray}\label{u13}
U_{13}=\left(
\begin{array}{ccc}
 \cos\theta               & 0 & \sin\theta{e^{-i\psi}} \\
     0                    & 1 &            0 \\
 -\sin\theta{e^{i\psi}} & 0 &        \cos\theta
\end{array}
\right),  
\end{eqnarray}
and $m_{1,2,3}^2$ are the light neutrino mass eigenvalues. Here the rotation angle $\theta$ and phase $\psi$ can be evaluated 
using the complex parameters in Eq. \eqref{mat:hgen}. From Eqs. \eqref{mat:hgen}-\eqref{mat:hpara} it is clear that there 
exists several parameters in $\mathcal{M}$ (obtained from the effective light neutrino matrix) to constrain $\theta$ and $\psi$ 
satisfying correct neutrino oscillation data. Therefore due to presence of several non-trivial matrices having many complex 
parameters in the effective  light neutrino mass matrix, it does not lead to very specific constraints on the parameters appearing in the neutrino linear seesaw mass matrix. 

It turns out, there is a way to have a more constrained scenario. Now along with the symmetry mentioned in Table \ref{tab:lss2}, for simplicity one can introduce an additional $Z_2$ symmetry
under which both $\eta$ and $\rho$ are odd (with all other particles are even under this symmetry). Therefore this two flavons 
will always appear together and under this additional symmetry the Lagrangian presented in Eq. \eqref{eq:lssnu} can be 
re-written in a simplified form as 
\begin{align}\label{eq:lsslag2}
 \mathcal{L}_{\nu} =
 & Y_{\nu N}\bar{L}\tilde{H}N_R+ \frac{Y_{RN}}{\Lambda}  \bar{\nu_{\R}}{N_L} 
\eta^{\dagger}\rho^{\dagger} + Y_{\nu S} \bar{L}\tilde{H}S_R+ \frac{Y'_{\nu 
S}}{\Lambda}\bar{S_L} \nu_R\eta\rho \nonumber \\ 
&~~~+ \bar{S_R} N_L (y_x\xi+ y_s\phi_S+y_a\phi_S)+ \bar{S_L} N_R 
(y'_x\xi^{\dagger}+ y'_s\phi^{\dagger}_S+y'_a\phi^{\dagger}_S)+\text{h.c.}.  
\end{align}
Subsequently, in the present set-up we work with this $Z_2$ symmetry to keep the analysis minimal and more predictive. Clearly, the $Y_{\nu N}$ and  
$Y_{\nu S}$ couplings remain unchanged and hence corresponding mass are given by Eq. \eqref{mat:mix1}. As the triplet 
flavon $\phi_s$ (and singlet $\xi$) do not share same $Z_2$ symmetry with $\eta$, the mass matrices involved in $S_L$-$\nu_{\R}$
and $\nu_{\R}$-$N_L$ couplings now can be written in much simpler way as 
\begin{eqnarray}\label{mat:mix2}
M'_{\nu S} = 
\frac{Y'_{\nu S}}{\Lambda} v_{\eta} v_{\rho} \mathbf{I},~~~~
m'_{\nu N} =\frac{Y_{RN}}{\Lambda} v_{\eta} 
v_{\rho} \mathbf{I}. 
\end{eqnarray}
In this simplified scenario, the mixing between the heavy neutrinos  $S_L - N_R$ and $S_R- N_L$ now takes the form 
\begin{eqnarray}\label{mat:heavy2}
 M'_{ N S} =\left(
\begin{array}{ccc}
 x_1 & 0   & s_1+a_1 \\
 0  & x_1 & 0\\
 s_1 - a_1 & 0 & x_1
\end{array}
\right),  M_{ N S} =\left(
\begin{array}{ccc}
 x_2 & 0   & s_2+a_2 \\
 0  & x_2 & 0\\
 s_2 - a_2 & 0 & x_2
\end{array}
\right)
\end{eqnarray}
where $x_1=y_{\xi_1} v_{\xi}$, $s_1=y_{s_1} v_{\s}, a_1=y_{a_1}  v_{\s}$, $x_2=y_{\xi_2} v_{\xi}$, $s_2=y_{s_2} v_{\s}$ and 
$a_2=y_{a_2}  v_{\s}$. Clearly, presence of the same $Z_2$ symmetry forbids any contribution from the singlet flavon $\eta$ in these 
matrices as evident from Eq. \eqref{eq:lsslag2}. Now, in this simplified scenario, substituting these mass matrices given in Eqs. 
\eqref{mat:mix1}, \eqref{mat:mix2} and \eqref{mat:heavy2} in the linear seesaw formula given in Eq. \eqref{eq:nulss} one can obtain  
the effective light neutrino mass matrix as 
{\small 
\begin{eqnarray}
 m_{\nu} &=& Y_{\nu N}v\frac{Y'_{\nu S}}{\Lambda} v_{\eta} v_{\rho}\left(
\begin{array}{ccc}
  x_1 & 0   & s_1+a_1 \\
 0  & x_1 & 0\\
 s_1 - a_1 & 0 & x_1
\end{array}
\right)^{-1}+Y_{\nu S}v\frac{Y_{RN}}{\Lambda} v_{\eta} 
v_{\rho}\left(
\begin{array}{ccc}
 x_2 & 0   & s_2+a_2 \\
 0  & x_2 & 0\\
 s_2 - a_2 & 0 & x_2
\end{array}
\right)^{-1}\nonumber\\
&=&\lambda_{1}\left(
\begin{array}{ccc}
 x_1 & 0   & -(a_1+s_1) \\
 0   & \frac{a_1^2-s_1^2+x_1^2}{x_1} & 0\\
 a_1-s_1     & 0 & x_1
\end{array}
\right)+\lambda_{2}\left(
\begin{array}{ccc}
 x_2 & 0   & -(a_2+s_2) \\
 0   & \frac{a_2^2-s_2^2+x_2^2}{x_2} & 0\\
 a_2-s_2     & 0 & x_2
\end{array}
\right),
\label{mnu:gen2}
\end{eqnarray}
}where $\lambda_1=\frac{Y_{\nu N}Y'_{\nu S}v v_{\eta}  v_{\rho}}{\Lambda(a_1^2-s_1^2+x_1^2)}$ and $\lambda_2=\frac{Y_{\nu S} 
Y_{RN}v  v_{\eta}v_{\rho}}{\Lambda(a_2^2-s_2^2+x_2^2)}$ are dimensionless  quantities.  Clearly, in the present scenario the
matrices involved in the heavy neutrino  mixing ($ M_{ N S}$ and  $M'_{ N S}$) dictate the pattern of light  neutrino mixing
as all other matrices are diagonal here. Furthermore, as mentioned  earlier, the hierarchy among the different mass matrices
is governed by the  specific discrete symmetries in order to ensure light neutrino mass of correct order.  The general 
structure for the light neutrino mass matrix originated from Dirac   linear seesaw as given in Eq. \eqref{mnu:gen2}, can be 
further analysed to  satisfy correct neutrino oscillation data. This in turn puts constraints on the  parameters appearing in
the neutrino matrix given in Eq. (\ref{mnu:gen2}).  Besides this, using these constrains on the complex mass parameters, 
one can  easily find the predictions involving neutrino mixing  angles, Dirac CP phase  and absolute masses for light neutrinos.
This predictive nature of the present  model makes it more interesting from the point of view of ongoing and upcoming  neutrino
experiments. Here we perform the analysis regarding the predictions for neutrino masses and mixing in two different frameworks.
First, in a simplest  scenario (Case A), we consider some equality between two terms (involving $A_4$ symmetric and anti-symmetric
contributions) appearing in the linear seesaw formula. Next, in a more general scenario (Case B), we do not consider any equality
among the symmetric and anti-symmetric terms in the effective light mass obtained linear seesaw formula and try to fit neutrino
oscillation data. We discuss these two cases below.

\subsection{Case A:}
In this simplest scenario, we first consider $\lambda_1=\lambda_2=\lambda$, $a_1=a_2=a, s_1=s_2=s$ and $x_1=x_2=x$. Hence the general 
structure for the effective light neutrino matrix as given in Eq. \eqref{mnu:gen2} reduces to 
\begin{eqnarray}
 m_{\nu}=2\lambda \left(
\begin{array}{ccc}
 x & 0   & -(a+s) \\
 0   & \frac{a^2-s^2+x^2}{x} & 0\\
 a-s     & 0 & x
\end{array}
\right).
\end{eqnarray}
Here $s$ and $a$ take care of the symmetric and anti-symmetric  contributions respectively originating from the two terms in the
linear seesaw formula. In order to diagonalise this mass matrix, let us first define a Hermitian  matrix as 
{\small 
\begin{align}\label{Mnu:simple}
 \mathcal{M}&=m_{\nu}m_{\nu}^{\dagger}\nonumber\\
        &=4|\lambda|^2 \left(
\begin{array}{ccc}
 |x|^2+|s+a|^2 & 0   & x(a-s)^*-x^*(a+s) \\
 0   & \frac{a^2-s^2+x^2}{x}\frac{(a^2-s^2+x^2)^*}{x^*}& 0\\
 x^*(a-s)-x(a+s)^*& 0 & |x|^2+|a-s|^2
\end{array}
\right).
\end{align}
}
This matrix, being Hermitian, can be diagonalised by a unitary matrix $U_{13}$, as given in  Eq. \eqref{u13} through the relation
$U_{13}^{\dagger}\mathcal{M}U_{13}={\rm diag}  (m_1^2,m_2^2,m_3^2)$. Here we find the mass eigenvalues ($m_1^2,m_2^2, m_3^2$) 
to be
\begin{eqnarray}
m_1^2&=&\kappa^2\left[1+\alpha^2+\beta^2-\sqrt{(2\alpha\beta\cos(\phi_{
ax}-\phi_{ sx}))^2+4(\alpha^2\sin^2\phi_{ax}+\beta^2\cos^2\phi_{sx}) }
\right],\label{eq:tm1}
\end{eqnarray}
\begin{eqnarray}
 m_2^2&=&\kappa^2\left[1+\alpha^4+\beta^4+2\alpha^2\cos 2 \phi_{ax} 
 -2\beta^2\cos 2 \phi_{sx}-2\alpha^2\beta^2\cos 2 (\phi_{sx} - \phi_{ax} 
)\right],\label{eq:tm2}\\
 m_3^2&=&\kappa^2\left[1+\alpha^2+\beta^2+\sqrt{(2\alpha\beta\cos(\phi_{
ax}-\phi_{ sx}))^2+4(\alpha^2\sin^2\phi_{ax}+\beta^2\cos^2\phi_{sx}) }
\right].\label{eq:tm3}
\end{eqnarray}
Here we have defined $\kappa^2=4|\lambda|^2|x|^2$, $\alpha=|a|/|x|$,  $\beta=|s|/|x|$, $\phi_{sx}=\phi_s - \phi_x $,
$\phi_{ax}=\phi_a - \phi_x$ with   $s=|s|e^{i\phi_s}$, $a=|a|e^{i\phi_a}$ and $x=|x|e^{i\phi_x}$ respectively.  For notational
convenience, the relative phases $\phi_{sx}$ and  $\phi_{ax}$ will be denoted just as $\phi_{s}$ and $\phi_{a}$ respectively 
from here onwards. It can be clearly seen from the expressions for mass eigenvalues that $m^2_3>m^2_1$ implying the preference 
for normal hierarchical light neutrino masses. From these  definitions it is clear that  $\alpha$ is associated with the 
anti-symmetric  contribution whereas $\beta$ is related to the symmetric contribution in the  Dirac neutrino mass matrix. Using
Eq. \eqref{eq:omega}, \eqref{u13}, the final  lepton mixing matrix in our  framework is given by 
\begin{eqnarray}\label{unu}
 U&=&U^{\dagger}_{\omega}U_{13}. 
\end{eqnarray}
Now, using Eq. \eqref{Mnu:simple} and Eq. \eqref{unu}, one can obtain the  correlation between the rotation angle $\theta$ and
phase $\psi$ as 
\begin{eqnarray}\label{eq:ang}
 \tan 2\theta=\frac{\beta\sin\phi_{s}\cos\psi-\alpha\cos\phi_{a}\sin\psi)}
 {\alpha\beta\cos(\phi_{s}-\phi_{a})}~~~ {\rm and}~~~ 
\tan\psi=-\frac{\alpha\sin\phi_{a}}
 {\beta\cos\phi_{s}}. 
\end{eqnarray}  
To extract the neutrino mixing angles in terms of the model parameters, we  compare this with the standard parametrisation of 
leptonic mixing matrix  known as Pontecorvo Maki Nakagawa Sakata (PMNS) mixing matrix given by 
\begin{equation}
U_{\text{PMNS}}=\left(\begin{array}{ccc}
c_{12}c_{13}& s_{12}c_{13}& s_{13}e^{-i\delta}\\
-s_{12}c_{23}-c_{12}s_{23}s_{13}e^{i\delta}& 
c_{12}c_{23}-s_{12}s_{23}s_{13}e^{i\delta} & s_{23}c_{13} \\
s_{12}s_{23}-c_{12}c_{23}s_{13}e^{i\delta} & 
-c_{12}s_{23}-s_{12}c_{23}s_{13}e^{i\delta}& c_{23}c_{13}
\end{array}\right), 
\label{PMNS}
\end{equation}
\begin{figure}[h]
$$
\includegraphics[height=5cm]{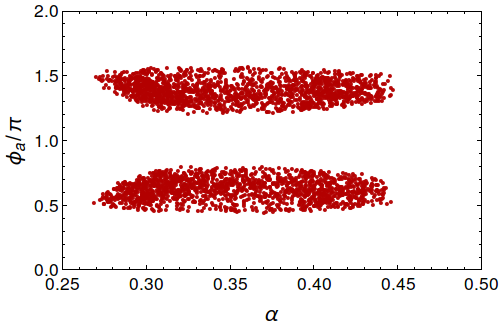}~~~~~~
\includegraphics[height=4.8cm]{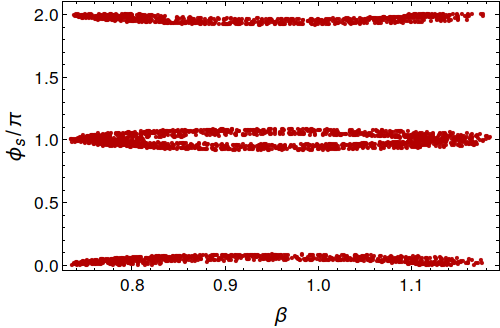}
$$
\caption{Allowed regions of $\alpha$-$\phi_a$ (left panel) and   $\beta$-$\phi_{s}$ (right panel) planes for 3$\sigma$ allowed
ranges of  $\theta_{13}$, $\theta_{12}$,  $\theta_{23}$ and the ratio ($r$) of solar to  atmospheric mass squared
differences~\cite{Esteban:2016qun, nufit}.}
\label{fig:abpp}
\end{figure}
and we obtain
\begin{eqnarray}
 \sin\theta_{13} e^{-i\delta}=\frac{1}{\sqrt{3}}(\cos\theta+\sin\theta 
e^{-i\psi}).
\end{eqnarray}
Now, $\sin\theta_{13}$ and $\delta$ can also be parametrised in terms of  $\theta$  and $\psi$ as 
\begin{eqnarray}\label{eq:s13}
 \sin^2\theta_{13}=\frac{1}{3}(1+\sin 2\theta\cos\psi)~~{\rm and}~~
 \tan\delta=\frac{\sin\theta\sin\psi}{\cos\theta+\sin\theta\cos\psi}.
\end{eqnarray}
Such correlation between the model parameters and neutrino mixing angles  $\theta_{13}, \theta_{12}, \theta_{23}$, Dirac CP phase
$\delta$ can also be  found in~\cite{Memenga:2013vc, Grimus:2008tt, Albright:2008rp,  Albright:2010ap, He:2011gb, Borah:2017dmk}.
Therefore from Eq. \eqref{eq:ang} and Eq. \eqref{eq:s13} it  is clear that the neutrino mixing angles are functions of four model
parameters namely, $\alpha$, $\beta$, $\phi_{s}$ and $\phi_{a}$. These are the parameters  associated with symmetric and 
anti-symmetric part of the effective light  neutrino mass matrix and corresponding relative phases. These parameters then 
can be constrained using the current data on neutrino mixing angles~\cite{Esteban:2016qun, deSalas:2017kay, nufit}. In addition 
to the bounds obtained from the mixing angles, the parameter space can be further constrained in oder to satisfy correct value 
for mass squared  differences. Here one can define a ratio for the solar to atmospheric mass  squared difference  as 
\begin{eqnarray}\label{eq:r}
r=\frac{\Delta{m}_{\odot}^{2}}{|\Delta{m}_{A}^{2}|}= 
\frac{\Delta{m_{21}^{2}}}{|\Delta{m}^2_{32}|}. 
\end{eqnarray}
\begin{figure}[h]
$$
\includegraphics[height=4.8cm]{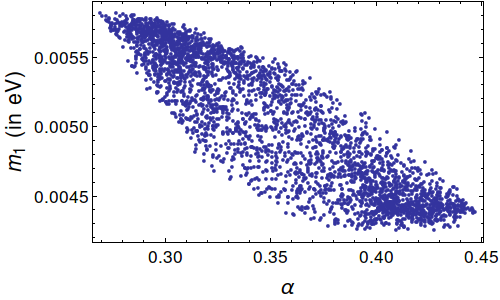}~~~~~~
\includegraphics[height=4.8cm]{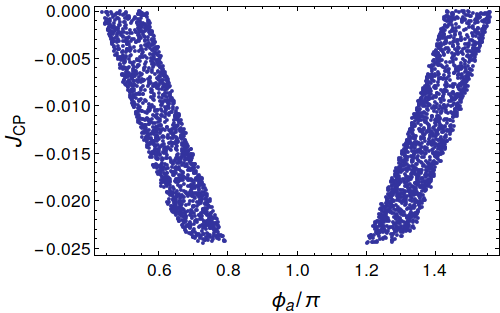}
$$
\caption{Predictions for lightest neutrino mass $m_1$ as a function of  $\alpha$ (left panel) and Jarlskog invariant $J_{CP}$
as a function of  $\phi_a$ (right panel). Here each points in both panels also satisfy 3$\sigma$  allowed ranges for  
$\theta_{13}$, $\theta_{12}$, $\theta_{23}$ and the ratio ($r$) of solar to  atmospheric mass squared 
differences~\cite{Esteban:2016qun,nufit}.}
\label{fig:m1jcs}
\end{figure}
From Eq. \eqref{eq:tm1}-\eqref{eq:tm3}, it is evident that this ratio $r$  is a function of the model parameters $\alpha$,
$\beta$, $\phi_{s}$ and $\phi_{a}$. In order  to satisfy correct neutrino oscillation data, we use the 3$\sigma$ allowed 
range of the  neutrino mixing angles and mass squared differences given in global fit analysis \cite{Esteban:2016qun,nufit}
to constrain these  model parameters. Here in Fig. \ref{fig:abpp} we have shown the allowed regions for parameters $\alpha$, 
$\beta$, $\phi_{s}$ and $\phi_{a}$ satisfying 3$\sigma$ ranges for neutrino mixing angles ($\theta_{13}, \theta_{12}, 
\theta_{23}$) and ratio of the mass squared differences $r$. In the left panel of Fig. \ref{fig:abpp} we show the allowed points 
in $\alpha$-$\phi_a$ plane whereas in the right panel we have plotted the same in $\beta$-$\phi_s$ plane. Here we find that the
parameter $\beta$,  associated with the symmetric part of the neutrino mass matrix ranges between 0.7-1.2 whereas the 
anti-symmetric part (contained in parameter $\alpha$) remains confined within a relatively narrower region
between 0.27 to 0.45. The associated phases also occupy distinct allowed regions as evident from both panels of Fig.  
\ref{fig:abpp}. After finding the allowed regions for $\alpha$, $\beta$, $\phi_{s}$ and $\phi_{a}$, one can easily find out the
common factor $\kappa$ appearing in the light neutrino mass eigenvalues using Eq. \eqref{eq:tm1}-\eqref{eq:tm2} and best fit
value of solar mass squared difference, $\Delta{m_{21}^{2}}=m^2_2-m^2_1=7.40\times 10^{-5}$ eV$^2$ \cite{Esteban:2016qun,nufit}.
The estimation of $\kappa$ then enables us to find predictions for absolute  neutrino masses. In the left panel of 
Fig. \ref{fig:m1jcs}, we plot the  predictions for lightest absolute neutrino mass $m_1$ as a function of $\alpha$  and it ranges
between ($0.42\times 10^{-2}$ -$0.58\times 10^{-2}$) eV for $\alpha$ in the range 0.27 to 0.45. Similarly, one can also find the 
estimates  for sum of all three absolute neutrino in this simplified scenario of the  neutrino mass matrix and is given by 
$\sum m_i=(0.062-0.070)$ eV, lying within the cosmological bound on sum of light neutrino masses $\sum  m_i  \leq 0.17$ eV from
Planck data~\cite{Ade:2015xua}. On the other hand, in the right panel of Fig. \ref{fig:m1jcs}, we have plotted  the allowed 
regions for the Jarlskog CP invariant $J_{\rm CP} = {\rm Im}[U_{e1}U_{\mu 2} U^*_{e2} U^*_{\mu 1}]$~\cite{Jarlskog:1985ht}  as a 
function of the  relative phase $\phi_a$ associated with the anti-symmetric contribution of the neutrino mass matrix and 
estimated to be within the range $|J_{CP}|\sim  0-0.024$. 
\begin{figure}[h]
$$
\includegraphics[height=4.8cm]{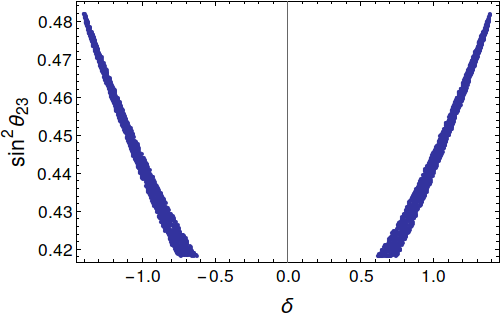}
$$
\caption{Predicted correlation between Dirac CP phase $\delta$ and  atmospheric mixing angle $\theta_{23}$ for Case A.}
\label{fig:d23s}
\end{figure}
In Fig. \ref{fig:d23s}, we show the most important  among such correlations namely, the one between the Dirac CP phase $\delta$ 
and atmospheric mixing angle $\theta_{23}$. Interestingly, here we find that, the model predicts the  CP  phase $\delta$ to be
in the range $-\pi/2\lesssim \delta \lesssim -\pi/5$ and  $\pi/5\lesssim \delta \lesssim \pi/2$ whereas $\sin^2\theta_{23}$ lies
in the  lower octant. This value of $\delta$ falls in the current preferred  ballpark suggested by experiments~\cite{Abe:2017uxa}
as well as global fit analysis \cite{Esteban:2016qun,nufit}, predicting atmospheric mixing angle $\theta_{23}$ to be in the lower
octant. 

\subsection{ Case B:}
In this subsection, we analyse the effective light neutrino mass matrix given in Eq. \eqref{mnu:gen} in a more general canvas to 
illustrate the effects of contributions  coming from symmetric and anti-symmetric parts appearing in the two different terms of 
the linear seesaw formula, without assuming any equality between two symmetric (and anti-symmetric) terms. Considering the most 
general structure for the light neutrino mass matrix as given in Eq. \eqref{mnu:gen}, we can define a Hermitian matrix as, 
\begin{align}
 \mathcal{M}&=m_{\nu}m_{\nu}^{\dagger}\\
        &=|\lambda|^2 \left(
\begin{array}{ccc}
 X_1 & 0   & X_2 \\
 0   & X_3 & 0\\
 X_4& 0 & X_5
\end{array}
\right)
\end{align}
where 
\begin{align}
 X_1&=4|x|^2+|(a_1+a_2)+(s_1+s_2)|^2,\nonumber\\
 X_2&=2x\{(a_1+a_2)^*-(s_1+s_2)^*\}-2x^*\{(a_1+a_2)+(s_1+s_2)\},\nonumber\\
 X_3&=\frac{1}{|x|^2}|(a_1^2+a_2^2)-(s_1^2+s_2^2)+2x^2|^2,\nonumber\\
 X_4&=2x^*\{(a_1+a_2)-(s_1+s_2)\}-2x\{(a_1+a_2)^*+(s_1+s_2)^*\},\nonumber\\
 X_5&=4|x|^2+|(a_1+a_2)-(s_1+s_2)|^2.\nonumber
\end{align}
Here for simplicity, we have considered  $x_1=x_2=x$  and  $\lambda_1=\lambda_2=\lambda$ while keeping the other terms distinct.
This Hermitian matrix $\mathcal{M}$ now can also  be diagonalised by a similar rotation matrix (in the 13 plane) given in  Eq. 
\eqref{u13} with rotation angle $\theta$ and phase factor $\psi$.  These parameters can therefore be expressed as 
\begin{eqnarray}\label{eq:u1parag}
 \tan 
2\theta&=&\frac{2\left[A\sin\psi- 
B\cos\psi\right]} 
{C_1+C_2+C_3+C_4},~~~
\tan\psi=-\frac{A}
 {B},
\end{eqnarray}
with 
\begin{eqnarray} 
A&=&(\alpha_1\sin\phi_{a_1}+\alpha_2\sin\phi_{a_2}),~~~B=(\beta_1\cos\phi_{
s_1}+\beta_2\cos\phi_{s_2}),\label{eq:A}\\
 C_1&=&\alpha_1\beta_1\cos(\phi_{a_1}-\phi_{s_1}),~~~~
 C_2=\alpha_1\beta_2\cos(\phi_{a_1}-\phi_{s_2}),\\
 C_3&=&\alpha_2\beta_1\cos(\phi_{a_2}-\phi_{s_1}),~~~~
 C_4=\alpha_2\beta_2\cos(\phi_{a_2}-\phi_{s_2})\label{eq:c4},
\end{eqnarray}
where we have defined the parameters as $\alpha_j=|a_j|/|x|$,  $\beta_j=|s_j|/|x|$,  $\phi_{a_jx}=\phi_{a_j} - \phi_x$,  
$\phi_{s_jx}=\phi_{s_j} - \phi_x $, $s_j=|s|e^{j\phi_{s_j}}$, $a_j=|a|e^{i\phi_{a_j}}$ and $x=|x|e^{i\phi_x}$ with $j=1,2$.
For notational  compactness we have written the relative phases $\phi_{a_jx}, \phi_{s_jx}$ in equation (\ref{eq:A}-\ref{eq:c4})
as $\phi_{a_j}, \phi_{s_j}$ with $j=1,2$. Hence, diagonalising the Hermitian matrix via  $U_{13}^{\dagger}\mathcal{M}U_{13}=
{\rm diag}(m_1^2,m_2^2,m_3^2)$, we obtain  the light neutrino masses as 
{\small 
\begin{eqnarray}
m_{1}^2&=&\kappa^2\left[
4+\alpha_1^2+\alpha_2^2+\beta_1^2+\beta_2^2+C_5  - 
\sqrt{4(C_1+C_2+C_3+C_4)^2+4^2(A^2+B^2) }
\right],\label{eq:gtm1}\\
m_2^2&=&\kappa^2\left[4+\alpha_1^4+\alpha_2^4+\beta_1^4+\beta_2^4+C_6\right],
\label{eq:gtm2}\\
m_{3}^2&=&\kappa^2\left[
4+\alpha_1^2+\alpha_2^2+\beta_1^2+\beta_2^2+C_5  + 
\sqrt{4(C_1+C_2+C_3+C_4)^2+4^2(A^2+B^2) }
\right],\label{eq:gtm3}
\end{eqnarray}
}
where
{\small 
\begin{eqnarray}
 C_5&=&2\{\alpha_1\alpha_2\cos(\phi_{ a_1}-\phi_{a_2})
       +\beta_1\beta_2\cos(\phi_{s_1}-\phi_{s_2})\},\nonumber\\
C_6 & =&4(\alpha_1^2\cos2\phi_{a_1}+\alpha_2^2\cos2\phi_{a_2})
    -4(\beta_1^2\cos2\phi_{s_1}+\beta_2^2\cos2\phi_{s_2}) \nonumber\\
&& +2\alpha_1^2\alpha_2^2\cos2(\phi_{a_1}-\phi_{a_2})
  -2\alpha_1^2\beta_1^2\cos2(\phi_{a_1}-\phi_{s_1})
  -2\alpha_1^2\beta_2^2\cos2(\phi_{a_1}-\phi_{s_2})\nonumber\\
&&  -2\alpha_2^2\beta_1^2\cos2(\phi_{a_2}-\phi_{s_1})
  -2\alpha_2^2\beta_2^2\cos2(\phi_{a_1}-\phi_{s_2})
  +2\beta_1^2\beta_2^2\cos2(\phi_{s_1}-\phi_{s_2}). \nonumber
\end{eqnarray}
}
Considering the contributions from both charged lepton and neutrino sectors, the  complete lepton mixing matrix in this general
case also is given by
\begin{eqnarray}
 U&=&U^{\dagger}_{\omega}U_{13}. 
\end{eqnarray}
\begin{figure}[h]
$$
\includegraphics[height=4.8cm]{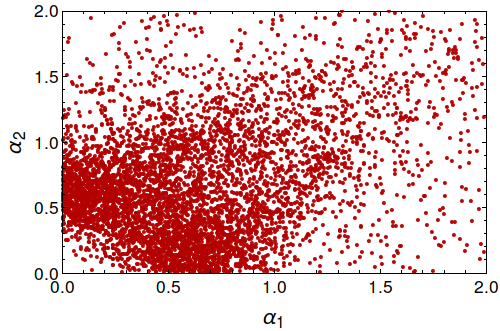}~
\includegraphics[height=4.8cm]{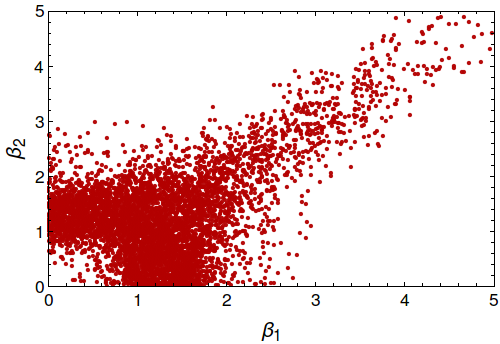}
$$
\caption{Allowed regions of $\alpha_1$-$\alpha_2$ (left panel) and   $\beta_1$-$\beta_2$ (right panel) planes for 3$\sigma$ 
allowed ranges of  $\theta_{13}$, $\theta_{12}$,  $\theta_{23}$ and the ratio ($r$) of solar to  atmospheric mass squared
differences~\cite{Esteban:2016qun,nufit}. These points additionally also satisfy the upper limit for sum of the three absolute 
neutrino masses $\sum m_i  \leq 0.17$ eV \cite{Abe:2017uxa}.}
\label{fig:a1a2b1b2}
\end{figure}

\begin{figure}[h]
$$
\includegraphics[height=4.8cm]{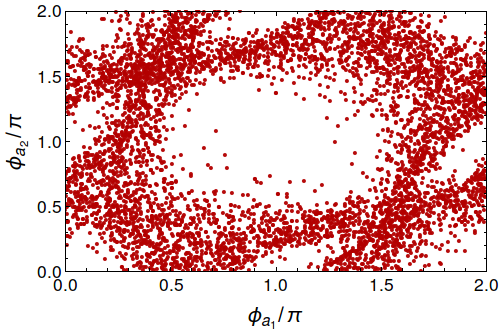}~
\includegraphics[height=4.8cm]{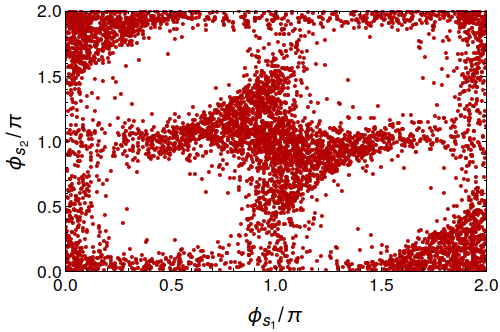}
$$
\caption{Allowed regions of $\phi_{a_1}$-$\phi_{a_2}$ (left panel) and   $\phi_{b_1}$-$\phi_{b_2}$ (right panel) planes for 
3$\sigma$ allowed ranges of  $\theta_{13}$, $\theta_{12}$,  $\theta_{23}$ and the ratio ($r$) of solar to  atmospheric mass
squared differences~\cite{Esteban:2016qun,nufit}. These points additionally also  satisfy the upper limit for sum of the three
absolute neutrino mass $\sum m_i  \leq 0.17$ eV \cite{Abe:2017uxa}.}
\label{fig:phg}
\end{figure}
Comparing this mixing matrix with $U_{\rm PMNS}$ as given in Eq. \eqref{PMNS}, one can obtain the correlations between the the
mixing angles ($\theta_{13}$,  $\theta_{12}$ and $\theta_{23}$) and Dirac CP phase $\delta$ as previously  given in 
Eq. \eqref{eq:s13}. Further using Eq. \eqref{eq:u1parag} we find the correspondence between neutrino mixing angles  and the 
relevant model parameters. Here the parameters $\alpha_j,  \beta_j, \phi_{a_j}$ and  $\phi_{s_j}$ with $j=1,2$ essentially 
dictate the  neutrino mixing patterns. Obviously, the number of parameters controlling neutrino mixing in this general case is 
more than what it was in the simple scenario described earlier. 
\begin{figure}[h]
$$
\includegraphics[height=4.8cm]{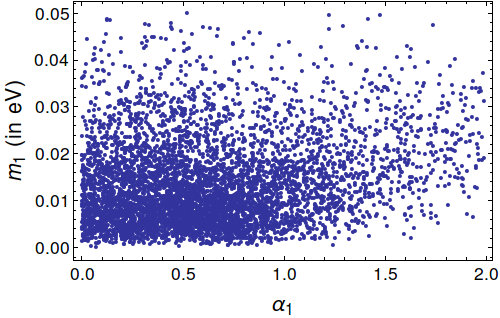}~
\includegraphics[height=4.8cm]{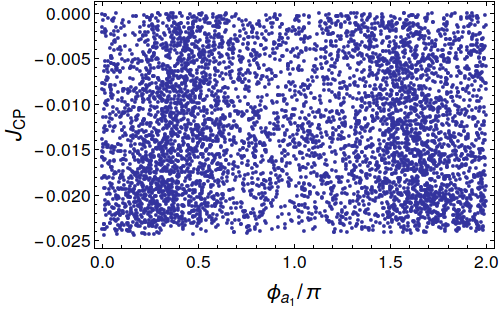}
$$
\caption{Predictions for lightest neutrino mass $m_1$ as a function of  $\alpha_1$ (left panel) and Jarlskog invariant
$J_{\rm CP}$ as a function of  $\phi_{a_1}$ (right panel). Here each points in both panels also satisfy 
3$\sigma$ allowed values of $\theta_{13}$, $\theta_{12}$, $\theta_{23}$ and the ratio ($r$) of  solar to 
atmospheric mass squared differences~\cite{Esteban:2016qun,nufit}.}
\label{fig:m1jg}
\end{figure}
Using Eq. \eqref{eq:gtm1}-\eqref{eq:gtm3} we define a ratio  $r(=\Delta{m}_{\odot}^{2}/|\Delta{m}_{A}^{2}|= 
\Delta{m_{21}^{2}}/|\Delta{m}^2_{32}|)$ in terms of very same parameters  $\alpha_j, \beta_j, \phi_{a_j}$ and  $\phi_{s_j}$. 
Here also  to satisfy  correct neutrino oscillation data, we use the 3$\sigma$ range of the  neutrino mixing angle and mass 
squared differences \cite{Esteban:2016qun,nufit} to constrain these parameters and we find the correlations among them. In
addition to the bounds from neutrino oscillation experiments, these parameters can also get   constrained in order to satisfy
the cosmological upper limit on sum of the three absolute  neutrino mass, given by $\sum m_i \leq 0.17$ eV~\cite{Abe:2017uxa}.
Therefore, using all  these constraints, in Fig. \ref{fig:a1a2b1b2} we have all the allowed points  in $\alpha_1$-$\alpha_2$ 
(left panel) plane and  $\beta_1$-$\beta_2$ (right  panel) plane respectively. In the left panel, we find that the  contributions
involving the anti-symmetric parts ($\alpha_1$ and $\alpha_2$) are  mostly confined within 0-2. On the other hand, as it is
evident from the right panel of  Fig. \ref{fig:a1a2b1b2}, the contributions involving the symmetric parts  ($\beta_1$ and
$\beta_2$) take relatively larger values satisfying correct neutrino oscillation data. Then in Fig. \ref{fig:phg}, we 
show the allowed parameter space in the  $\phi_{a_1}$-$\phi_{a_2}$ plane (left panel) and $\phi_{b_1}$-$\phi_{b_2}$ 
plane (right panel) respectively which clearly show distinct correlations between relative phases associated with 
symmetric and anti-symmetric contributions. 
\begin{figure}[h]
$$
\includegraphics[height=4.8cm]{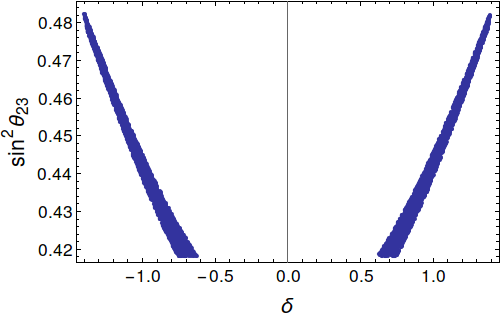}
$$
\caption{Predicted correlation between Dirac CP phase $\delta$ and  atmospheric mixing angle $\theta_{23}$ for Case B.}
\label{fig:d23g}
\end{figure}
After finding the allowed regions for $\alpha_{1,2}$, $\beta_{1,2}$,  $\phi_{a_{1,2}}$ and $\phi_{s_{1,2}}$, one can again find
out the common  factor $\kappa$ appearing in the light neutrino mass eigenvalues involved  in this general case using 
Eq. \eqref{eq:gtm1}-\eqref{eq:gtm2} and best  fit value of solar mass squared difference, $\Delta{m_{21}^{2}} = m^2_2-m^2_1 = 
7.40\times 10^{-5}$ eV$^2$~\cite{Esteban:2016qun,nufit} as mentioned earlier. The estimate for  $\kappa$ enables us to find 
predictions for absolute neutrino masses using  Eq. \eqref{eq:gtm1}-\eqref{eq:gtm2}. In the left panel of Fig. \ref{fig:m1jg},
we show the predictions for lightest absolute neutrino mass $m_1$ as a  function of $\alpha_1$ (parameter involved in one of the
anti-symmetric  contribution). It can be seen from this plot that $m_1$ can be as large as $0.05$ eV for $\alpha_1$ within the 
limit of 2. Such values of the lightest neutrino mass correspond to sum of all three absolute  neutrino masses $\sum 
m_i=(0.06-0.17)$ eV saturating the cosmological upper limit.  In the right panel of Fig. \ref{fig:m1jg}, we have 
again plotted the allowed regions for the Jarlskog CP invariant $J_{\rm CP}$ as a  function of the relative phase $\phi_{a_1}$ 
associated with one of the  anti-symmetric contribution of the neutrino mass matrix and estimated to be  within the range 
$|J_{\rm CP}|\sim 0-0.024$, analogous to the previous result. In Fig. \ref{fig:d23g}, we have now plotted the correlation
between the Dirac CP  phase $\delta$ and atmospheric mixing angle $\theta_{23}$. Similar to the  previous case, here also we 
find that, the model predicts the Dirac CP phase  $\delta$ to be in the range $-\pi/2\lesssim \delta \lesssim -\pi/5$ and 
$\pi/5\lesssim \delta \lesssim \pi/2$ whereas $\theta_{23}$ lies in the  lower octant. 

In this two different limits of Dirac linear seesaw discussed above, we have observed that the allowed range of the light 
neutrino mass is different in the two cases.  In Case A, due to much constrained scenario of the neutrino mass matrix 
\begin{figure}[h]
$$
\includegraphics[height=4.8cm]{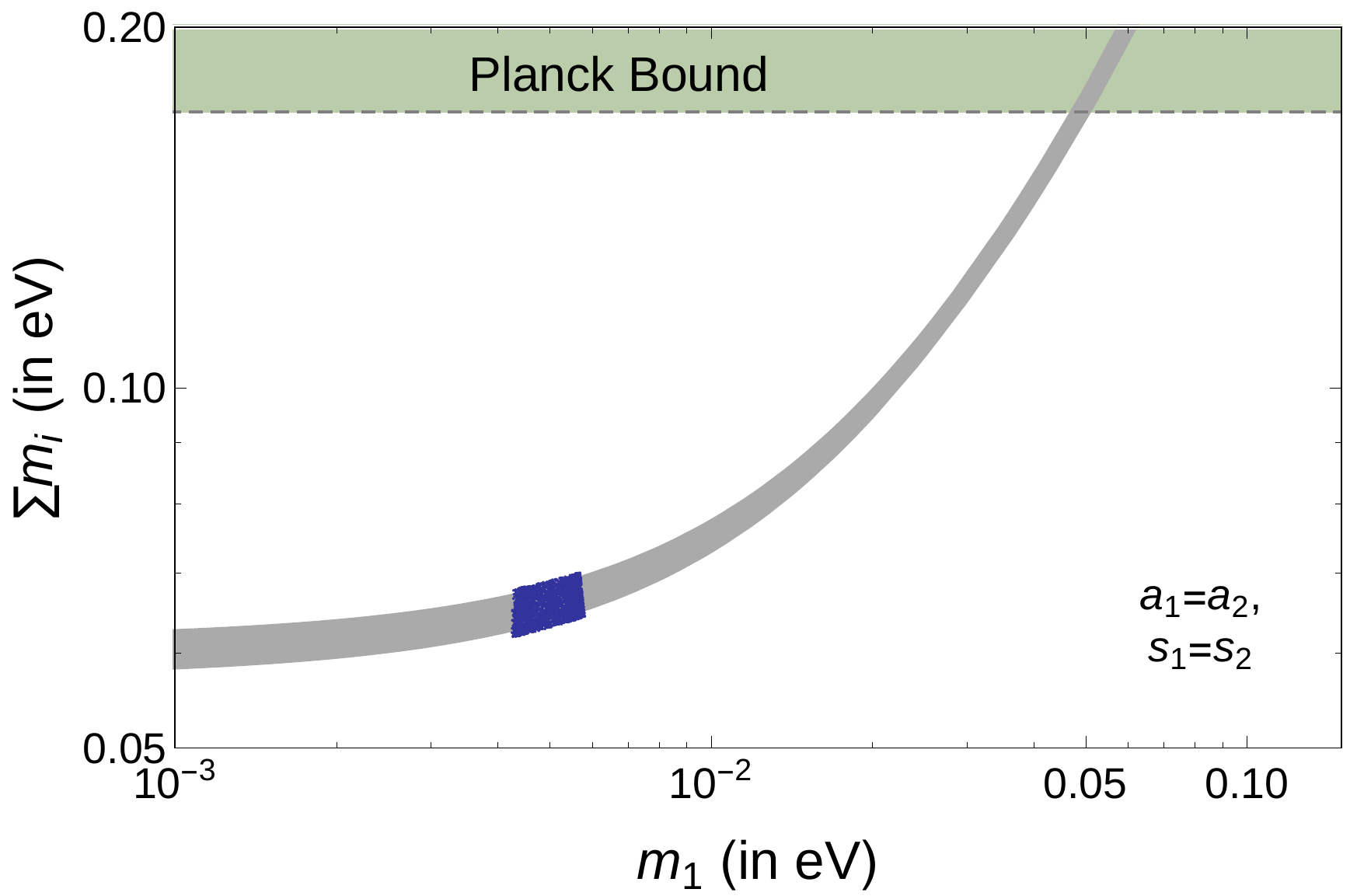}~~~~~~
\includegraphics[height=4.8cm]{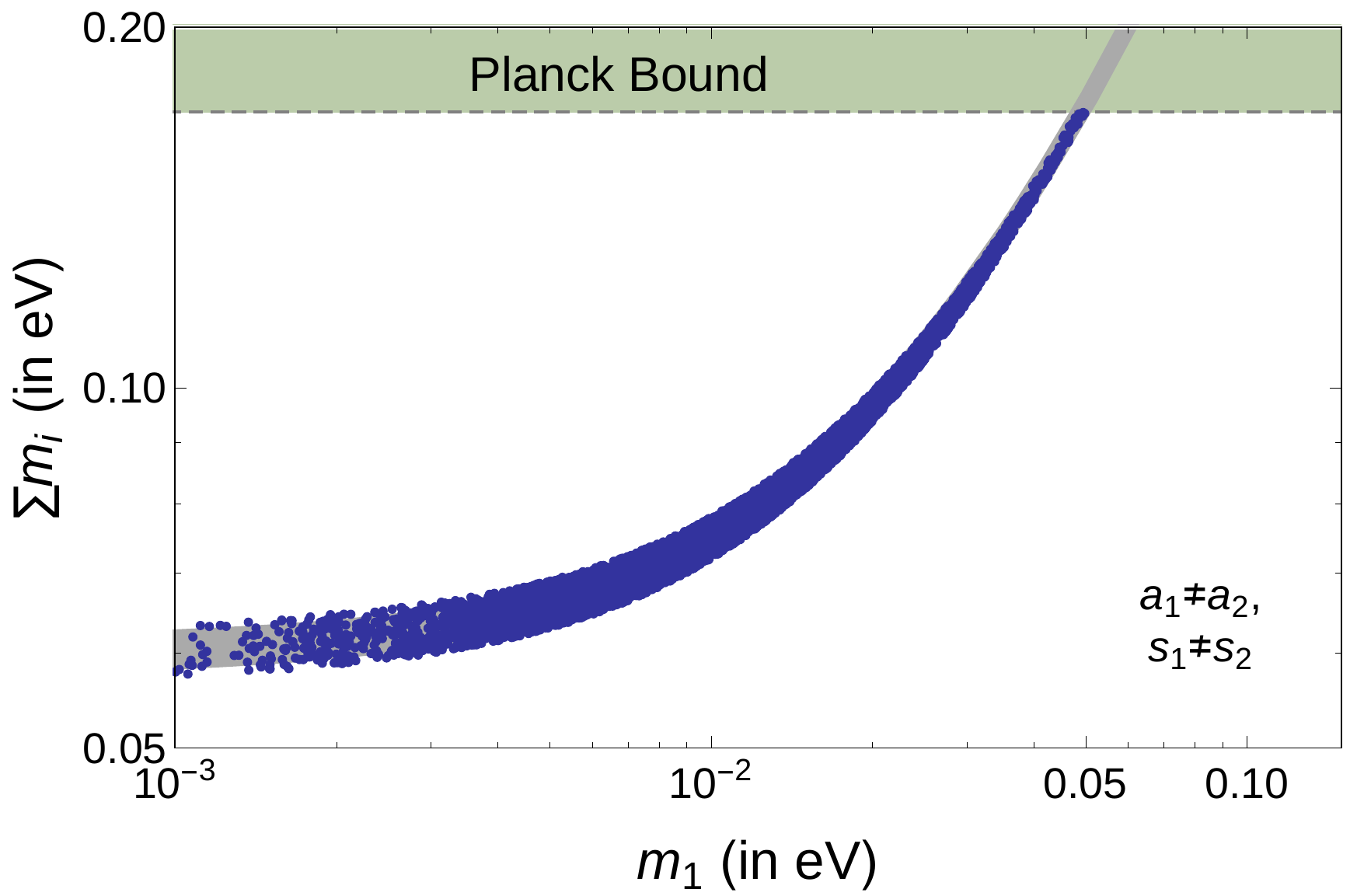}
$$
\caption{Comparison for $\sum m_i$ as a function of lightest neutrino mass $m_1$ in both cases A and B.}
\label{fig:mcomp}
\end{figure}
the correlation between $m_1$  and $\sum m_i$ is mainly concentrated within  a narrow region. This is shown in the left panel of
Fig. \ref{fig:mcomp}.   However, for the Case B, due to presence of more number of parameters originated  from both the 
contributions of linear seesaw formula being distinct, the constrains are much more relaxed. In this case neutrino masses can 
vary from a very small values to larger ones saturating the cosmological upper bound $\sum  m_i  \leq 0.17$ eV. Here the whole 
region is basically allowed, as shown in the right panel of Fig. \ref{fig:mcomp}. Finally, it is important to mention that,
within these two different limits, inverted hierarchy of neutrino mass is not allowed, another important prediction of our model.

\section{Conclusion}
\label{sec2}
We have proposed a linear seesaw model for Dirac neutrinos in this work. After proposing a renormalisable version of such seesaw by extending the SM with appropriate fields and global symmetries, we move on to a more predictive framework of $A_4$ flavour symmetry. While in the renormalisable version, the seesaw scale can be brought down to the TeV scale by assuming smallness of certain terms in the seesaw mass matrix, which can be justified by incorporating additional approximate global symmetries. In the $A_4$ symmetric version additional discrete and global lepton number symmetry are chosen in a manner to make sure  that the correct hierarchy between different terms 
appearing in the complete  neutral fermion mass matrix is naturally obtained without making any assumptions or referring to approximate global symmetries. The
interesting feature of the conventional linear seesaw framework where a small lepton number breaking term in seesaw formula,
linear in Dirac neutrino mass, can give rise to correct neutrino mass with heavy neutrinos lying in TeV scale, is retained in 
the Dirac version of it by appropriately generating hierarchical terms at different orders (dimension four and dimension five). 
Since lepton  number is present as a global unbroken symmetry in the model, all the mass matrices involved are of Dirac type and
hence the $A_4$ triple products contain the  anti-symmetric components which play a crucial role in  generating the correct 
neutrino phenomenology. Since we use the  $S$ diagonal basis of $A_4$ for Dirac neutrino case, the charged lepton mass  matrix is
also non trivial in our scenarios and hence can contribute to the  leptonic mixing matrix. For generic choices of $A_4$ flavon 
alignments, we find that the model remains very predictive in terms of neutrino mass hierarchy, leptonic CP phase, octant of
atmospheric mixing angle as well as absolute neutrino masses. While the neutrino mass hierarchy is predicted to be the normal one, the Dirac CP phase 
$\delta$ is found to lie in the range $-\pi/2\lesssim \delta \lesssim -\pi/5$ and  $\pi/5\lesssim \delta \lesssim \pi/2$ whereas
the atmospheric mixing angle $\theta_{23}$ lies in the lower octant. The predictions for lightest neutrino mass, in one of the 
scenarios, can saturate the cosmological upper bound on the sum of absolute neutrino masses $\sum  m_i  \leq 0.17$ eV. Apart from
these, being a model predicting Dirac neutrinos, it can also predict the absence of lepton number violation and hence can not be 
tested in ongoing and future neutrinoless double beta decay experiments. These aspects keep the detection prospects of the model
very promising at experiments ranging from neutrino oscillations, cosmology to rare decay ones.

\acknowledgments
One of the authors, DB acknowledges the hospitality and facilities provided by Seoul-Tech, Korea and KEK Theory Center, Japan 
during June, 2018 where part of this work was completed. 

\appendix
\section{$A_4$ Multiplication Rules}
\label{appen1}
$A_4$, the symmetry group of a tetrahedron, is a discrete non-abelian group of 
even permutations of four objects. It has four irreducible representations: 
three one-dimensional and one three dimensional which are denoted by $\bf{1}, 
\bf{1'}, \bf{1''}$ and $\bf{3}$ respectively, being consistent with the sum of 
square of the dimensions $\sum_i n_i^2=12$. We denote a generic permutation 
$(1,2,3,4) \rightarrow (n_1, n_2, n_3, n_4)$ simply by $(n_1 n_2 n_3 n_4)$. The 
group $A_4$ can be generated by two basic permutations $S$ and $T$ given by $S = 
(4321), T=(2314)$. This satisfies 
$$ S^2=T^3 =(ST)^3=1$$
which is called a presentation of the group. Their product rules of the 
irreducible representations are given as
$$ \bf{1} \otimes \bf{1} = \bf{1}$$
$$ \bf{1'}\otimes \bf{1'} = \bf{1''}$$
$$ \bf{1'} \otimes \bf{1''} = \bf{1} $$
$$ \bf{1''} \otimes \bf{1''} = \bf{1'}$$
$$ \bf{3} \otimes \bf{3} = \bf{1} \otimes \bf{1'} \otimes \bf{1''} \otimes 
\bf{3}_a \otimes \bf{3}_s $$
where $a$ and $s$ in the subscript corresponds to anti-symmetric and symmetric 
parts respectively. Denoting two triplets as $(a_1, b_1, c_1)$ and $(a_2, b_2, 
c_2)$ respectively, their direct product can be decomposed into the direct sum 
mentioned above. In the $S$ diagonal basis, the products are given as
$$ \bf{1} \backsim a_1a_2+b_1b_2+c_1c_2$$
$$ \bf{1'} \backsim a_1 a_2 + \omega^2 b_1 b_2 + \omega c_1 c_2$$
$$ \bf{1''} \backsim a_1 a_2 + \omega b_1 b_2 + \omega^2 c_1 c_2$$
$$\bf{3}_s \backsim (b_1c_2+c_1b_2, c_1a_2+a_1c_2, a_1b_2+b_1a_2)$$
$$ \bf{3}_a \backsim (b_1c_2-c_1b_2, c_1a_2-a_1c_2, a_1b_2-b_1a_2)$$
In the $T$ diagonal basis on the other hand, they can be written as
$$ \bf{1} \backsim a_1a_2+b_1c_2+c_1b_2$$
$$ \bf{1'} \backsim c_1c_2+a_1b_2+b_1a_2$$
$$ \bf{1''} \backsim b_1b_2+c_1a_2+a_1c_2$$
$$\bf{3}_s \backsim \frac{1}{3}(2a_1a_2-b_1c_2-c_1b_2, 2c_1c_2-a_1b_2-b_1a_2, 
2b_1b_2-a_1c_2-c_1a_2)$$
$$ \bf{3}_a \backsim \frac{1}{2}(b_1c_2-c_1b_2, a_1b_2-b_1a_2, c_1a_2-a_1c_2)$$

\bibliographystyle{apsrev}
\bibliography{ref_fimp.bib}

\end{document}